\documentclass[english,twocolumn,superscriptaddress,
  longbibliography]{revtex4-1}

\usepackage{soul}
\usepackage[dvipsnames]{xcolor}

\usepackage{xcolor}
\usepackage{graphicx}
\usepackage{amsmath}
\usepackage{amsfonts}
\usepackage{color}
\usepackage{babel}
\usepackage{url}
\usepackage{hyperref}
\usepackage{tikz}
\usepackage{CJKutf8}
\usepackage[title]{appendix}
\usepackage{siunitx}

\definecolor{lime}{HTML}{A6CE39}
\DeclareRobustCommand{\orcidicon}{
	\begin{tikzpicture}
	\draw[lime, fill=lime] (0,0) 
	circle [radius=0.16] 
	node[white] {{\fontfamily{qag}\selectfont \tiny ID}};
	\draw[white, fill=white] (-0.0625,0.095) 
	circle [radius=0.007];
	\end{tikzpicture}
	\hspace{-2mm}
}

\foreach \x in {A, ..., Z}{\expandafter\xdef\csname orcid\x\endcsname{\noexpand\href{https://orcid.org/\csname orcidauthor\x\endcsname} {\noexpand\orcidicon}} }

\begin{document}

\title{Construction of approximate invariants for non-integrable Hamiltonian systems}

\author{Yongjun Li\orcidA{}}\thanks{email: yli@bnl.gov}
\affiliation{Brookhaven National Laboratory, Upton, New York 11973, USA}
\author{Derong Xu}
\affiliation{Brookhaven National Laboratory, Upton, New York 11973, USA}
\author{Yue Hao}\affiliation{Michigan State University, East Lansing, Michigan  48864, USA}

\begin{abstract}
  We present a method to construct high-order polynomial approximate invariants (AI) for non-integrable Hamiltonian dynamical systems, and apply it to a modern ring-based particle accelerator. Taking advantage of a special property of one-turn transformation maps in the form of a square matrix, AIs can be constructed order-by-order iteratively. Evaluating AI with simulation data, we observe that AI's fluctuation is actually a measure of chaos. Through minimizing the fluctuations, the stable region of long-term motions, i.e., the dynamic aperture of the accelerator, could be enlarged.
\end{abstract}
 
\maketitle

\section{Introduction}
 The construction of the invariant phase-space tori of Hamiltonian systems is important in many mathematical physics fields (see ref.~\cite{hagel1986invariants,warnock1991close,kaasalainen1994construction,kaasalainen1995construction,celletti2000improved,laakso2013canonical,yu2017,belanger2025} and references therein). For non-integrable Hamiltonian systems, exact solutions for motion are often impossible. They generally lack sufficient conserved quantities. However, Kolmogorov-Arnold-Moser theory~\cite{de2001tutorial} proves that approximate invariants (AI) could still exist in such systems. The theory addresses the stability of motion in near-integrable Hamiltonian systems that can be seen as small perturbations of integrable systems. If the perturbation is small enough, a significant portion of the invariant tori of the integrable system survive, though deformed. These surviving tori act as quasi-invariants or AIs, where the motion remains confined and quasi-periodic, preserving the structure of the integrable case. Even though resonances can disrupt some tori, many tori persist in a ``cantorized" form, as long as certain non-resonance conditions are met. The remnants of the AI tori provide a structure that partially resists chaotic diffusion, preserving long-term stability in certain regions of phase space. Such region is also known as ``dynamic aperture (DA)" in particle accelerator physics~\cite{chao2023handbook}. A sufficient DA is often required for beam manipulation and machine operation. It is important to explicitly obtain the AIs of given lattices and evaluate their fluctuations in designing and optimizing modern accelerator magnetic lattice~\cite{hagel1986invariants,warnock1991close}. 

\section{Method}
\begin{figure}
    \centering
    \includegraphics[width=0.95\columnwidth]{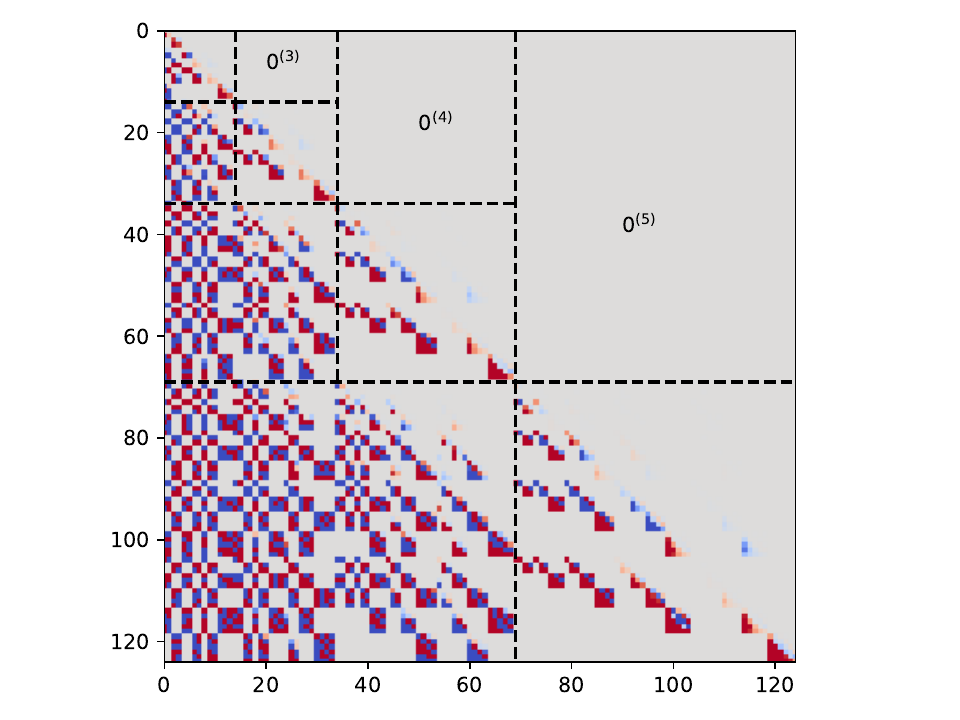}
    \caption{Up to $5^{th}$-order blocked square matrix $M^T$. The first top-left $14\times14$ block is used to construct linear AIs $W^{(2)}_{1,2}$. Then the expanded $34\times34$, $69\times69$ and $125\times125, \cdots$ square matrices are used to obtain high order AIs $W^{(3,4,5,\cdots)}_{1,2}$ respectively, and so on. Note that all elements in the top-right block's ($0^{(3-5)}$) are 0's.}
    \label{fig:block_square}
 \end{figure}

 The evolution of motion in a periodical Hamiltonian system can be approximately represented with polynomial maps. These maps can be extracted with the truncated power series algorithm~\cite{berz1995modern,yang2009array,zhang2024cpptpsa} to any arbitrary order. For example, consider a 4-dimensional phase space vector $[x,p_x,y,p_y]^T$, which represents an on-momentum charged particle's motion in the transverse plane in an accelerator. The superscript $^T$ represents the transpose of matrices or vectors. First we construct a $\Omega$-th order vector
 \begin{equation}
    Z = [x,p_x,y,p_y,x^2,xp_x,xy,xp_y,p^2_x,p_xy,\cdots,p^{\Omega}_y]^T.
 \end{equation}
 Let the one-turn map of the system be written in a matrix form as
 \begin{equation}
    Z_{final}=MZ_{initial},
 \end{equation}
 here $M$ is a square matrix. Such map is usually non-symplectic due to the truncation of high order polynomial terms. Following the method in Chao's note~\cite{chao2020lectures}, an $\Omega$-order Taylor series approximate invariant $W$ can be written as
 \begin{equation}
    W^{(\Omega)} = V^TZ,
 \end{equation}
 with $V$ as the coefficient vector yet to be found. Since $W$ is an invariant, for any $Z$ after one-turn transformation through $M$, we have
 \begin{equation}
    V^TZ=V^TMZ \Longrightarrow V=M^TV.
 \end{equation}
 It means that the vector $V$ needs to be $M^T$'s eigenvector with its eigenvalue as 1. However, when applying a direct eigen-analysis on the $\Omega(\ge3)$-order matrix $M^T$, it usually fails to yield proper AIs. It is because the coefficients of higher orders in $M^T$ are significantly greater than the lower orders. More specifically, the magnitude incremental step at each $n^{th}$ order is $n!$. Therefore, the highest order terms dominate the eigen-analysis, and the obtained eigen-vectors, usually have near to zero coefficients in the low order terms. It might yield multiple but independent eigen-vectors. To overcome this difficulty, we introduce a method to start with the well-known linear Wronskian invariants (aka Courant-Snyder invariants~\cite{courant1958} for uncoupled linear motion) and then iteratively construct higher order AIs by taking advantage of the special property of the square matrix. In this paper, the National Synchrotron Light Source II (NSLS-II) storage ring~\cite{dierker2007} is used as a real-world example for illustration. Its $5^{th}$ one-turn square matrix $M$ is constructed by sequentially concatenated the magnetic elements as shown in Fig.~\ref{fig:block_square}. $M^T$ matrix is a sparse matrix with its elements in the top-right blocks are zeros, because any $n^{th}$ order term in $Z$ must be a function of the same and even-higher orders terms ($\geq n^{th}$), but not of the lower order terms ($<n^{th}$), i.e.,
 \begin{equation}
     Z_{final}^n = \sum_{i\geq n} m_i Z_{initial}^i
 \end{equation}
 This property actually enables us to construct AIs order-by-order iteratively. 

 We start with the linear invariants, then construct higher order AIs on top of them. Even in the linear case, $Z$ must be kept up to the quadratic terms in order to find two quadratic invariants $W^{(2)}_{1,2}$. Here superscript $^{(2)}$ indicates the order of $W$, and subscripts $_{1,2}$ are their indexes. First, a $14\times14$ square matrix $M^T$ truncated up to the quadratic terms is used. The eigen-analysis on $M^T$ yields two eigen-vectors $V^{(2)}_{1,2}$ with their eigenvalues as 1: 
 \begin{verbatim}
          V2_1         |          V2_2
 ----------------------|----------------------
 2 0 0 0  4.883952e-02 | 0 0 2 0  2.969949e-01
 0 2 0 0  2.047522e+01 | 0 0 0 2  3.367061e+00
 \end{verbatim}
 Here each monomial is represented with the first four integers as the power indexes of $x,p_x,y,p_y$, and the last float as its coefficient. These two invariants can be easily recognized as the Courant-Synder invariants in the horizontal and vertical planes respectively,
 $W_{u=x,y}=\gamma_u u^2+2\alpha_u up_u+\beta_u p_u^2$, with the Twiss parameters $\alpha,\beta,\gamma$. The disappearance of two crossing terms $xp_x$ and $yp_y$ terms is simply due to the local $\alpha_{x,y}=0$.

 Next we construct cubic AIs by adding some additional cubic terms into linear invariants $V=[V^{(2)},V^{(3)}]$, here $V^{(3)}$ represents the cubic coefficients yet to be found. By observing the form of the cubic square matrix $M^T\,(34\times34)$ in Fig.~\ref{fig:block_square}, we notice that its upright blocks $0^{(3)}$ are composed of zeros, therefore, it can be divided into 4 sub-matrices $M^T = 
     \begin{bmatrix}
        m_0 & 0^{(3)} \\
        m_2 & m_3
     \end{bmatrix}$.
 Here $m_0$ with a dimension $(14\times14)$ is the quadratic matrix already being analyzed previously, $0^{(3)}\;(14\times20)$ is a zero matrix, $m_2\;(20\times14)$ and $m_3\;(20\times20)$ are two matrices with the cubic term coefficients. The new cubic AIs need to satisfy the following condition:
 \begin{equation}\label{eq:block}
    \begin{bmatrix}
        V^{(2)} \\
        V^{(3)}
    \end{bmatrix}
  = 
    \begin{bmatrix}
        m_0 & 0^{(3)} \\
        m_2 & m_3
    \end{bmatrix}
    \begin{bmatrix}
        V^{(2)} \\
        V^{(3)}
    \end{bmatrix}.
 \end{equation}
 Because the first row in Eq.~\eqref{eq:block} is already valid automatically, the cubic term coefficient $V^{(3)}$ can be simply solved with:
 \begin{equation}
    V^{(3)} = (I-m_3)^{-1}m_2V^{(2)}.
    \label{eq:iter}
 \end{equation}

\begin{figure}
    \centering
    \includegraphics[width=0.95\columnwidth]{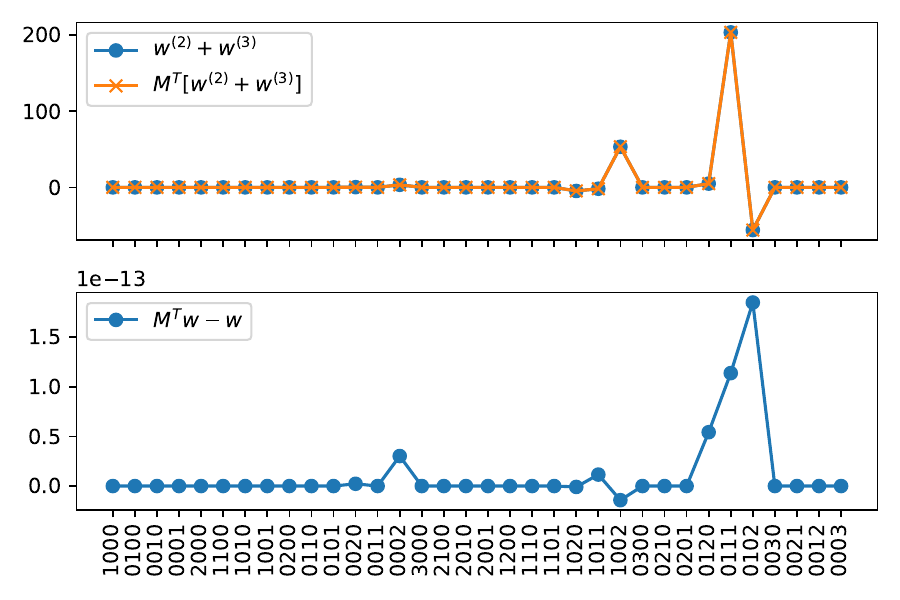}
    \caption{Comparison a $3^{rd}$-order AI's coefficient $[V^{(2)}+V^{(3)}]$ with its one-turn transformation $M^T[V^{(2)}+V^{(3)}]$. Top: the values of all 34 monomial coefficients. Bottom: the difference between them represents the quality of the AI.}
    \label{fig:cubic_error2}
 \end{figure}

 By solving Eq.~\eqref{eq:iter}, the cubic terms in two AIs are obtained as:
 \begin{verbatim}
          V3_1         |          V3_2
 ----------------------|----------------------
 3 0 0 0  1.327349e-01 | 1 0 2 0 -4.703005e+00
 2 1 0 0 -6.078392e-01 | 1 0 1 1 -1.778960e+00
 ...                   | ...
 0 1 0 2  4.370392e+01 | 0 1 0 2 -5.597472e+01
 \end{verbatim}
 The constructed cubic AI's coefficients $[V^{(2)},V^{(3)}]^T$ can be compared with its one-turn transformation $M^T[V^{(2)},V^{(3)}]^T$ as shown in Fig.~\ref{fig:cubic_error2}. The maximum difference of coefficients is around $3\times10^{-13}$. It is worthwhile to mention that, the existence of the solution in Eq.~\eqref{eq:iter} requires the square matrix $(I-m_3)$ to be invertible. In some cases, this condition might not be satisfied, which means the appearance of broken tori ~\cite{chao2020lectures} due to a $3^{rd}$ resonance. However, for most stable systems such as particle accelerators, their $(I-m_3)$ usually invertible because their fundamental frequency is intentionally chosen to be off-resonant. Once the cubic AIs are obtained, we can iteratively apply Eq.~\eqref{eq:block} to obtain even higher order AIs.

\section{Application in lattice design}
\begin{figure}
    \centering
    \includegraphics[width=0.9\linewidth]{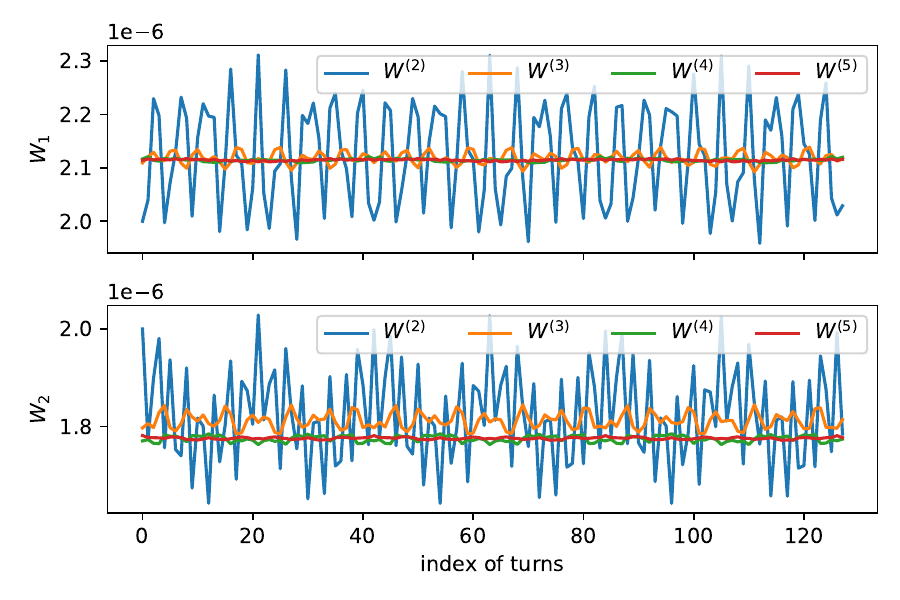}
    \caption{Fluctuations of two AIs $W^{(2,3,4,5)}_{1,2}$ at different orders ($2^{nd}-5^{th}$) by evaluating with a simulated trajectory.}
    \label{fig:trackInv5}
 \end{figure}

 Once the AIs are obtained, we can verify them with simulation data. Here a simulated 128 turns trajectory starting with an initial condition $x=6.4\si{mm},\,y=2.6\si{mm},\,p_{x,y}=0$ is obtained with a symplectic tracking algorithm~\cite{yoshida1990construction}. After evaluating different order AIs $W^{(2,3,4,5)}_{1,2}$ with the simulated data, they are observed converging gradually with reduced fluctuations as illustrated in Fig.~\ref{fig:trackInv5}, which indicates the accuracy of AIs is improved by including more higher order terms. However, the AIs' quality becomes worse when their order $\Omega$ is above 5 in this specific example. Thus far, it is unclear if the higher-order AIs are spoiled by the accumulated numerical errors, or they simply don't exist at all.

\begin{figure}
    \centering
    \includegraphics[width=0.95\columnwidth]{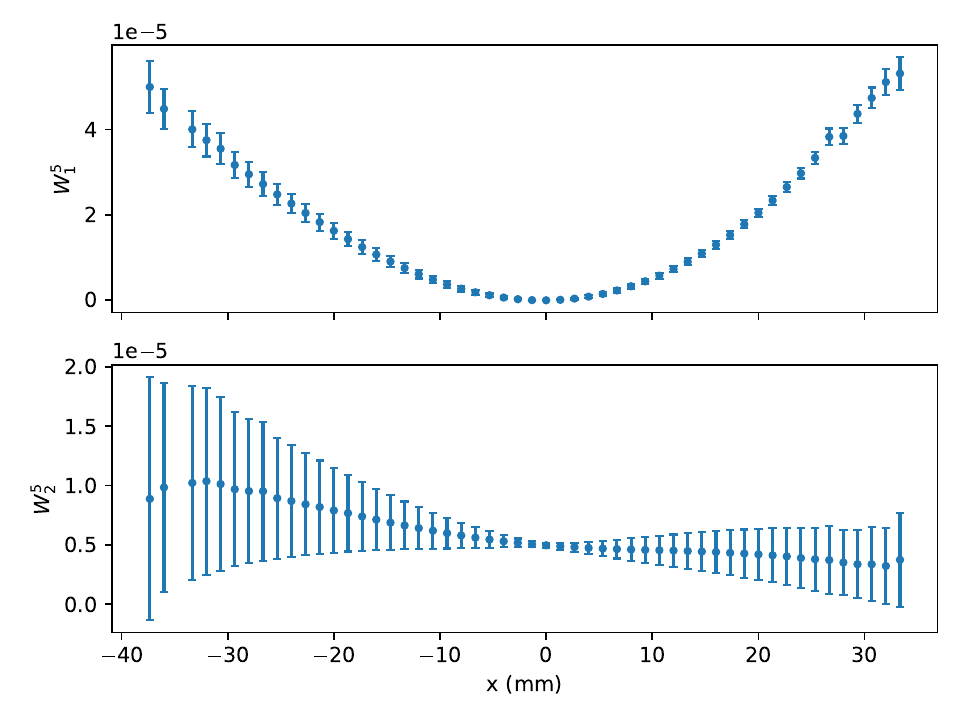}
    \caption{Absolute fluctuations of the $5^{th}$-order AIs with the horizontal oscillation amplitude (the initial vertical amplitude is $6\,\si{mm}$) in the NSLS-II ring.}
    \label{fig:ai_vs_amp}
 \end{figure}

 As seen in Fig.~\ref{fig:trackInv5}, these AIs are only approximately constant, rather than exactly. After evaluating with simulated data, AI's fluctuation increase gradually with oscillation amplitude as shown in Fig.~\ref{fig:ai_vs_amp}. The fluctuations of linear AIs (aka ``the action smear") are traditionally used as a chaos indicator in accelerator physics~\cite{chao1988}. Once the fluctuations exceed a certain range, approximate tori are broken and the stable particle motion can not be maintained.
 
\begin{figure}
    \centering
    \includegraphics[width=0.95\columnwidth]{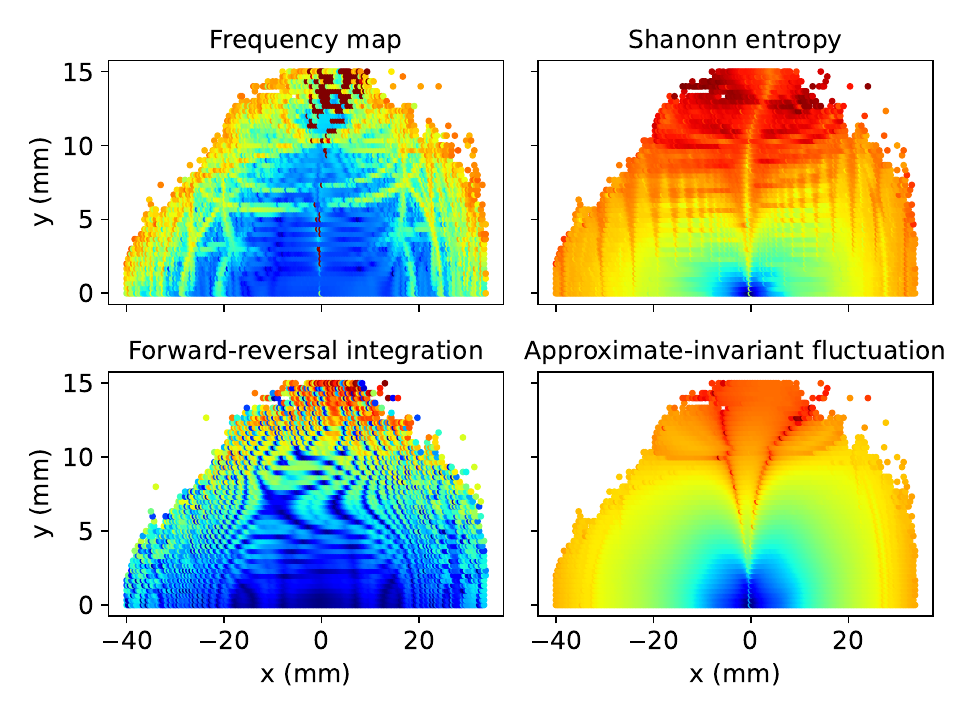}
    \caption{Comparison of different chaos indicators for the NSLS-II ring. Top-left: frequency map analysis; top-right: Shannon entropy; bottom-left: forward-reversal integration; bottom-right: fluctuations of approximate invariant. They display a common pattern, in which regular motions gradually dilute into chaos with the amplitude.}
    \label{fig:fluctation5}
 \end{figure}

 There exist many other chaos indicators (some of them are compared in Ref.~\cite{bazzani2023,montanari2025chaos}), which can be used for studying nonlinear beam dynamics. Among them, the frequency map analysis (FMA)~\cite{laskar2003,papaphilippou2014}, Shannon entropy~\cite{li2025}, forward-reversal integration~\cite{hwang2020,li2021,panichi2017}, are selected to compare with the fluctuation of AIs for the NSLS-II ring in Fig.~\ref{fig:fluctation5}. Here the relative AIs' fluctuation is defined as $\frac{\sigma_{W_1}}{|\left<W_1\right>|}+\frac{\sigma_{W_2}}{|\left<W_2\right>|}$. $\sigma_{W_{1,2}}$ represent the standard deviations of multiple-turn $W$'s, and $|\left<W_{1,2}\right>|$ are their absolute mean values. In Fig.~\ref{fig:fluctation5}, various chaos indicators display a common pattern, in which regular motions gradually dilute into chaos with amplitude. Note that the color scales used in these 4 cases are not directly-comparable, because of the different chaos metrics used. From the comparison, we observe that the FMA excels at detecting resonances, which needs a spectral analysis, such as FFT or NAFF. Such spectral analysis is also adopted in ref.~\cite{yu2017}. The forward-reversal integration mainly describes the unpredictability of long-term motions, relying on the randomness in computation. The Shannon entropy and AI's fluctuations are actually similar. Both describe the  fluctuations of predefined motion constants. While computing Shannon entropy, the phase space region needs to be discretized into a mesh grid, which makes the transition from regular motion to chaos not as smooth as AI. While AI has an explicit function form, the characterization of chaos is more accurate.
 
\begin{figure}
    \centering
    \includegraphics[width=0.95\columnwidth]{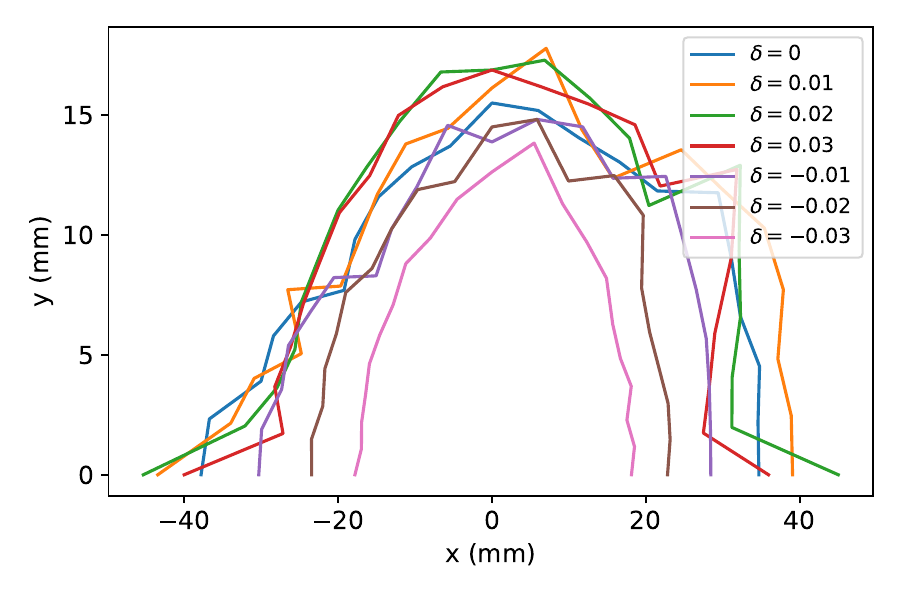}
    \caption{On- and off-momentum dynamic apertures for the optimal solution obtained by minimizing the fluctuations of AIs. Here $\delta=\frac{\Delta P}{P}$ represents charged particle's relative momentum deviations.}
    \label{fig:da_opt}
 \end{figure}

 In designing accelerators, it is generally believed that, by suppressing the chaos of beam motions, their dynamic apertures can be enlarged. Since AI's fluctuations is a measure of chaos, we explore the possibility of optimizing the NSLS-II magnetic lattice by suppressing the fluctuations of its $5^{th}$ order AIs. Technically, by tuning some harmonic sextupoles, we aim to minimize the AI's fluctuations on a specific trajectory starting with a given initial condition $x=15.7\si{mm},\,y=3.7\si{mm},p_{x,y}=0$. Once the optimal sextupole configuration is reached, the on- and off-momentum dynamic apertures are obtained by tracking particles for 2048 turns with the simulation code \textsc{elegant}~\cite{borland2000elegant} and illustrated in Fig.~\ref{fig:da_opt}. More detailed simulations confirm that, this nonlinear lattice performance is similar to our baseline design~\cite{dierker2007}. However, the suppression of AI's fluctuation is more efficient than other multi-particle tracking-based optimizations, such as ~\cite{yang2011multiobjective,li2018}. The optimization is accomplished within several hours even on a single core computer.    
 
 It is worth to mention another method to construct approximate actions also from a square matrix in the ref.~\cite{yu2017}. Its square matrix is constructed but with the monomials of the phase space variables $z=\bar{x}-i\bar{p}$ after applying the Courant-Synder parameterization. Then a Jordan analysis is carried out to determine the amplitude-dependent-tune-shift and approximate actions $W_0$. Predicting the long-term stability of each initial condition in phase space requires an additional transformation from the approximate action-angle pair to a rigid rotation, achieved through an iterative process in the frequency domain. Mathematically, such Jordan decomposition is a block-diagonal matrix with Jordan blocks. Ref.~\cite{yu2017} further explains a method of detecting resonances, including high-order weak ones, which demands substantial computational resources such as high-dimensional Fourier analysis. Alternatively, our proposed method utilizes the eigen-analysis on the raw square matrix (without the Courant-Synder parameterization). It aims to finding a diagonal matrix (or as close to diagonal as possible) that is similar to the original matrix. The needed computational effort significantly reduces. However, the most valuable information, i.e., the approximate invariants, are constructed explicitly. With the obtained AIs, other tune related information, such as the tune-shift-with-amplitude, could be further derived using the methods explained in refs.~\cite{nagaitsev2020,mitchell2021}.

\section{Conclusion}
 In summary, we present a method to construct the approximate invariants of non-integrable Hamiltonian systems and demonstrate its application on a modern accelerator. We also demonstrate the fluctuation of AIs when evaluating with simulation data is a chaos indicator, which can be potentially used as an efficient optimization objective in designing modern accelerators.

\begin{acknowledgments}
 We would like to thank S. Nagaitsev (BNL), C. Mitchell (LBNL) for their valuable discussion and constructive suggestion, P. Belanger and G. Sterbini (CERN) for a detailed numerical comparison and confirmation, H. Zhang (JLab), Y. Hidaka and D. Hidas (BNL) for their consistent supports in computation. This research is supported by the U.S. Department of Energy (DOE) under Contract No. DE-SC0012704, Office of Basic Energy Sciences, Field Work Proposal 2025-BNL-PS040, and the DOE HEP award DE-SC0019403.

 The data that support the findings of this article are not publicly available because of legal restrictions preventing unrestricted public distribution. The data are available from the authors upon reasonable request.
\end{acknowledgments}

\bibliography{ref.bib}

\end{document}